\documentclass[11pt]{article}
\usepackage{geometry}    % See geometry.pdf to learn the layout options. There are lots.
\usepackage{latexsym}
\usepackage{graphics}
\usepackage{graphicx}
\usepackage{epstopdf}
\usepackage{epsfig}
\usepackage{amssymb}
\usepackage{makeidx}
\usepackage{amsfonts}
\usepackage{amstext}
\usepackage{amsmath}
\usepackage{amsbsy}
\usepackage{wasysym}
\usepackage{mathrsfs}
\usepackage{authblk}
\newlength{\InitialBLS}
\author[a,*]{A.A. Machado}
\author[b]{E. Segreto,}
\author[c]{D. Warner,}
\author[b]{A. Fauth,}
\author[b]{B. Gelli,}
\author[b]{R. M\'aximo,}
\author[b]{A. Pizolatti}
\author[a]{L. Paulucci,}
\author[d]{F. Marinho,}

\affil[a]{Universidade Federal do ABC (UFABC),\\ Av. dos Estados, 5001, Santo Andr\'e, SP, 09210-170, Brazil}
\affil[b]{Instituto de F\'isica Gleb Wataghin, Universidade Estadual de Campinas - Unicamp,\\
	Rua Sergio Buarque de Holanda, No 777, CEP 13083-859 Campinas, SP, Brazil}
\affil[c]{Colorado State University,\\ Fort Collins, Colorado 80523 USA}
\affil[d]{Universidade Federal de S\~ao Carlos, \\Rodovia Anhanguera, km 174, 13604-900, Araras, SP,
	Brazil}

\title{The X-ARAPUCA: An improvement of the ARAPUCA device}

\begin{document}
	\maketitle
	\let\oldthefootnote\thefootnote
	\renewcommand{\thefootnote}{\fnsymbol{footnote}}
	\footnotetext[1]{Corresponding author - E-mail: aameliabm@gmail.com}
	\let\thefootnote\oldthefootnote

\begin{abstract}

The ARAPUCA is a novel technology for the detection of liquid argon scintillation light, which has been proposed for the far detector of the Deep Underground Neutrino Experiment.
The X-ARAPUCA is an improvement to the original ARAPUCA design, retaining the original ARAPUCA concept of photon trapping inside a highly reflective box while using a wavelength shifting slab inside the box to increase the probability of collecting trapped photons onto a silicon photomultiplier array. The X-ARAPUCA concept is presented and its performances are compared to those of a standard ARAPUCA by means of analytical calculations and Monte Carlo simulations.   
\end{abstract}

%As for the standard ARAPUCA, light enters the detector through a dichroic window which has the property of being highly transparent for wavelengths below a certain cutoff, while being highly reflective above it. For the X-ARAPUCA, the filter plate is coated on the external surface with a TPB wavelength shifter, which converts the VUV 127 nm light (LAr Scintillation light) to the region where the filter is transparent.
%The novel concept for the X-ARAPUCA consists of adding a waveshifting light guide inside the box, in the form of a WLS plastic slab with approximately the same dimensions of the acceptance filter window.  Photons entering the ARAPUCA are wavelength shifted to a wavelength reflected by the filter, capturing them inside it.  These photons will be guided in two different ways by the SiPM array: They can be captured in the waveshifting light guide by total internal reflection, or be captured between the reflecting inner surfaces of the box and reach the array after a few reflections (as in a standard ARAPUCA). Simulations of this design suggest that this modification will lead to a substantial increase of the ARAPUCA collection efficiency.
%}

%\keywords{Liquid Argon, Scintillation, TPC, Photosensor.}

%\proceeding{LIDINE 2017: LIght Detection In Noble Elements\\
 % 22-24 September 2017\\
%SLAC National Accelerator Laboratory}

\section{Introduction}

The Deep Underground Neutrino Experiment (DUNE) \cite{DUNE} will represent one of the most relevant experimental efforts towards a comprehensive understanding of neutrino properties in the next two or three decades. The primary science objectives of the DUNE experiment are the investigation of CP violation in the leptonic sector, the determination of the ordering of neutrino masses and the precision tests of the three neutrinos paradigm.

DUNE foresees the realization of a neutrino beam and of a near detector, both located at the Fermi National Accelerator Laboratory (Fermilab), Illinois, and a far detector 
based on the technology of Liquid Argon (LAr) TPC that will be installed at the Sanford Underground Research Facility in South Dakota, located 1,300 km away. This is the ideal 
distance since it will give the highest sensitivity in measuring the CP violating phase and in determining the mass hierarchy \cite{DUNE2}.  Moreover the huge active mass of the far detector 
will allow to develop a rich program of non accelerator physics, that includes the search for proton decay, the detection of supernova neutrinos and of atmospheric neutrinos. 

LArTPC technique allows to perform an accurate 3-D and calorimetric
reconstruction of the ionizing events which happen inside its active volume 
and can benefit from the detection of LAr scintillation light. LAr is known to be an 
abundant scintillator, emitting about 40 photons/keV of energy deposited by 
minimum ionizing particles, in absence of electric fields. The light is emitted in the Vacuum Ultra Violet (VUV) region of the electromagnetic spectrum in a 10 nanometers band centered around 127 nm \cite{doke}. This 
circumstance complicates its detection, since 
common (cryogenic) photo-sensitive devices with glass or fused silica windows
are insensitive to these wavelengths.
The current paradigm for LAr scintillation light detection foresees the use 
of wavelength shifting (WLS) compounds which absorb VUV light and re-emits 
it in the visible, where it can be more easily detected.\\  
    
The X-ARAPUCA is a development of the ARAPUCA concept \cite{Machado2016} which is being considered as one of the possible alternatives  for LAr scintillation light detection in the DUNE far detector.

\section{Operating principle}

The X-ARAPUCA  is not only a development and an optimization of the 
traditional ARAPUCA one, but it is conceived as a mutation of the original 
idea and it represents a new perspective for the photon detection system of the Deep Underground Neutrino Experiment (DUNE).

X-ARAPUCA is a hybrid 
solution between an ARAPUCA \cite{Machado2016} and a light guide \cite{bars}.

%which should allow to increase its detection efficiency with respect to a standard ARAPUCA by minimizing the number of overall reflections on its internal surfaces 

A standard ARAPUCA consists of a box with highly reflective internal surfaces 
and acceptance windows constituted by a shortpass dichroic filter coated 
with two different WLS, one on each side. The external shifter converts the 
127 nm  photons produced by the scintillation of liquid argon into a photon 
with wavelength below the filter's cutoff, where it is assumed to be 
transparent, while the inner one shifts the light to a wavelength above the 
cutoff where the filter is reflective. The net result is that the
photon is trapped inside the highly reflective box and it can be detected by 
the active photo-sensors installed on its internal surface.\\ 

\begin{figure}[h]
\begin{center}
\includegraphics[width=15cm,angle=0]{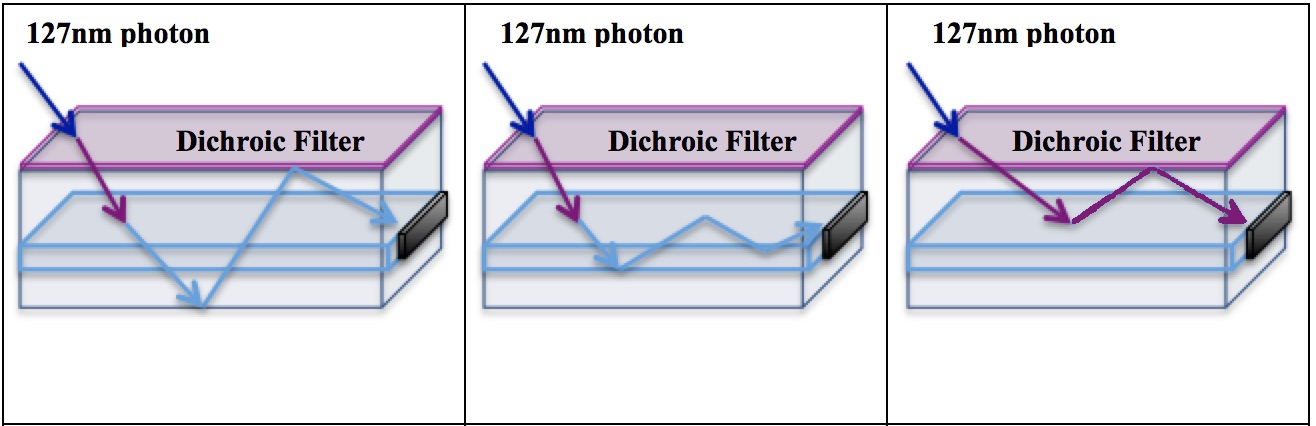}
\caption{Left: Standard ARAPUCA mechanism. The photon enters the box, it is converted by the WLS slab and is captured inside the ARAPUCA.
Center: Total internal reflection. A converted photon enters the box and it is converted by the WLS in the slab and trapped by total internal reflection. Right: High angle photons. A photon with high incidence angle inside the box, is trapped in the liquid argon gap between the filter and the acrylic slab. Notice that in this last case the guided photons are not shifted by the slab.}
\label{fig:threeways}
\end{center}
\end{figure}

In the case of the  X-ARAPUCA the inner shifter is substituted by a an
acrylic slab which has the WLS compound embedded. The active photo-sensors 
are optically coupled to one or more sides of the slab itself, as shown in 
figure \ref{fig:threeways}.
% In this way a fraction of the photons will be 
%converted inside the slab, trapped by total internal reflection and guided 
%to the array of photo-sensors. Photons which are not trapped by total 
%internal reflection can be detected through the standard ARAPUCA mechanism.

There are two main mechanisms through which a photon can be detected by the X-ARAPUCA:
\begin{enumerate}
	\item {\it Standard ARAPUCA mechanism.} The photon, after entering the X-ARAPUCA box, is converted by the WLS of the inner slab, but is not captured by total internal reflection. In this case the photon bounces a few times on the inner surfaces of the box until when it is or detected or absorbed (figure \ref{fig:threeways},left); 
	
	\item {\it Total internal reflection.} The photon, converted by the filter and the slab, gets trapped by total internal reflection. It will be guided towards one end of the slab where it will be eventually detected. This represents the first improvement with respect to a conventional ARAPUCA (figure \ref{fig:threeways},center), which contributes to reduce the effective number of reflections on the internal surfaces. The sides of the slab where there are not active photo-sensors will be coated with a reflective layer which will allow to keep the photon trapped by total internal reflection.
\end{enumerate}

In addition to these, a third mechanism which can guide the photons towards 
the photo-sensors can be envisaged. A photon impinging on the surface of the guiding slab with large angle of incidence will most likely be reflected (the refractive index of the liquid argon is about 1.23 \cite{cerenkov} and that of the acrylic slab is typically around 1.5). The reflected photon will return to the filter at the same large angle of incidence and again it will be, with high probability, reflected (the filter is made of fused silica and has a refractive index of about 1.5). Large angle photons are preferably guided within the first LAr gap to the side of the box where the photo-sensors will be installed, (figure \ref{fig:threeways},right)

\section{Expected collection efficiency}
\label{sec:analytical}
If the X-ARAPUCA were constructed with the same shifters and dichroic filter used for the ARAPUCA, the only difference would be the collection mechanism of the photons. The baseline option is to use a shortpass dichroic filter with a cut-off wavelength at 400 nm coated with para-Terphenyl (pTP) \cite{pTP} on the external side (emission wavelength $\simeq$ 350 nm) and TetraPhenyl-Butadiene (TPB) \cite{TPB} \cite{TPB_ettore} (emission wavelength $\simeq$ 430 nm) inside, or deposited directly on the filter or embedded in the acrylic slab in the case of X-ARAPUCA.\\
The fraction of photons which is able to cross the dichroic filter and which is converted by TPB depends on the properties of the filter and of the shifters and can be considered to be the same for the ARAPUCA and the X-ARAPUCA with good approximation. The main difference is that in one case the TPB is deposited by vacuum evaporation on the inner side of the filter and in the other it is embedded inside an acrylic slab. Small differences in conversion efficiencies of 350 nm photons to 430 nm ones are expected. In fact, in both cases, the TPB is directly excited by the 350 nm photon and no mechanisms of energy transfer are active, since the acrylic is highly transparent to this wavelength. Furthermore, waveshifting slabs with extremely high conversion efficiencies (> 90\%) are available on the market (see for example \cite{eljen}).

The collection efficiency of an ARAPUCA or of a X-ARAPUCA is the fraction of photons which is collected on the active photo-sensors with respect to the photons entering the box. The detection of the photons is not considered here, for this reason the characteristics of the SiPMs are not considered.
   
%The increase in efficiency of the X-ARAPUCA with respect to a standard ARAPUCA can be easily estimated analytically. For this purpose, consider a standard ARAPUCA configuration: a box made of a PTFE with reflectivity of 0.98 observed by one array of SiPMs with an area of 80$\times$4 mm$^2$  installed on one of the short sides of the box. 

The collection efficiency of an ARAPUCA can be calculated through the formula \cite{paper_ettore}:
\begin{equation}
\epsilon_{coll}^A=\frac{f}{1-R(1-f)}
\label{eq:eff_coll_arapuca}
\end{equation}

where {\it f} is the active coverage and R is the average reflectivity of the inner surfaces. 
%In the present case f=0.017 and R=0.98 (assuming that the dichroic filter has the same reflectivity to shifted photons as PTFE) and $\epsilon_{coll}^{ARAPUCA}$=0.41

In the case of the X-ARAPUCA a TPB acrylic slab is installed in the middle of the box and it is optically coupled to a SiPM array. The conversion of the photons is assumed to happen inside the slab and a fraction of them is trapped by total internal reflection.
 The critical angle for total internal reflection inside the slab is given by:

\begin{equation}
\theta_{critical}=asin \frac{n_{acrylic}(430 nm)}{n_{LAr}(430 nm)} = asin \frac{1.23}{1.5} \simeq 55^o
\label{eq:critical_angle}
\end{equation}

%The solid angle delimited by the critical angle is given by:
%\begin{equation}
%\Omega=4 \pi cos(\theta_{critical})
%\end{equation}

The total fraction of photons trapped inside the slab is:

\begin{equation}
\epsilon_{trapped}=\frac{4 \pi cos(\theta_{critical})}{4 \pi} = cos(\theta_{critical}) = 0.57
\end{equation}

Trapped photons will bounce inside the slab until when they are eventually detected. The fraction of trapped photons which is collected by the SiPM can be calculated through equation \ref{eq:eff_coll_arapuca} considering that the refections can happen only on the sides of the slab where there are no SiPM. Reflections on these sides of the slab are assumed to be specular in order to keep the reflected photon trapped inside the slab itself. Defining {\it f$_{slab}$} as the fraction of the lateral surface of the slab covered by SiPMs, the collection efficiency of an X-ARAPUCA can be written as:
\begin{equation}
\epsilon_{coll}^X = (1-\epsilon_{trapped})\frac{f}{1-R(1-f)}+\epsilon_{trapped}\frac{f_{slab}}{1-R(1-f_{slab})}
\label{eq:eff_coll_xarapuca}
\end{equation}

Under the following simplifying assumptions:
\begin{enumerate}
	\item Square ARAPUCA/X-ARAPUCA with side of length {\it L} and internal height {\it h}; 
	\item The SiPM array has the same height, {\it h}, of the inner box and its surface can be written as {\it q$\times$L$\times$h}, with {\it q} ranging from 0 to 4; 
\end{enumerate}

%equations \ref{eq:eff_coll_arapuca} and \ref{eq:eff_coll_xarapuca} can be written in term of 

{\it f} can be defined as:

\begin{equation}
f=q\frac{1}{2(L/h+2)}
\end{equation} 
 
and {\it f$_{slab}$} as:

\begin{equation}
f_{slab}= q/4
\end{equation} 
 
\begin{figure}[tbh]
	\includegraphics[width=7cm]{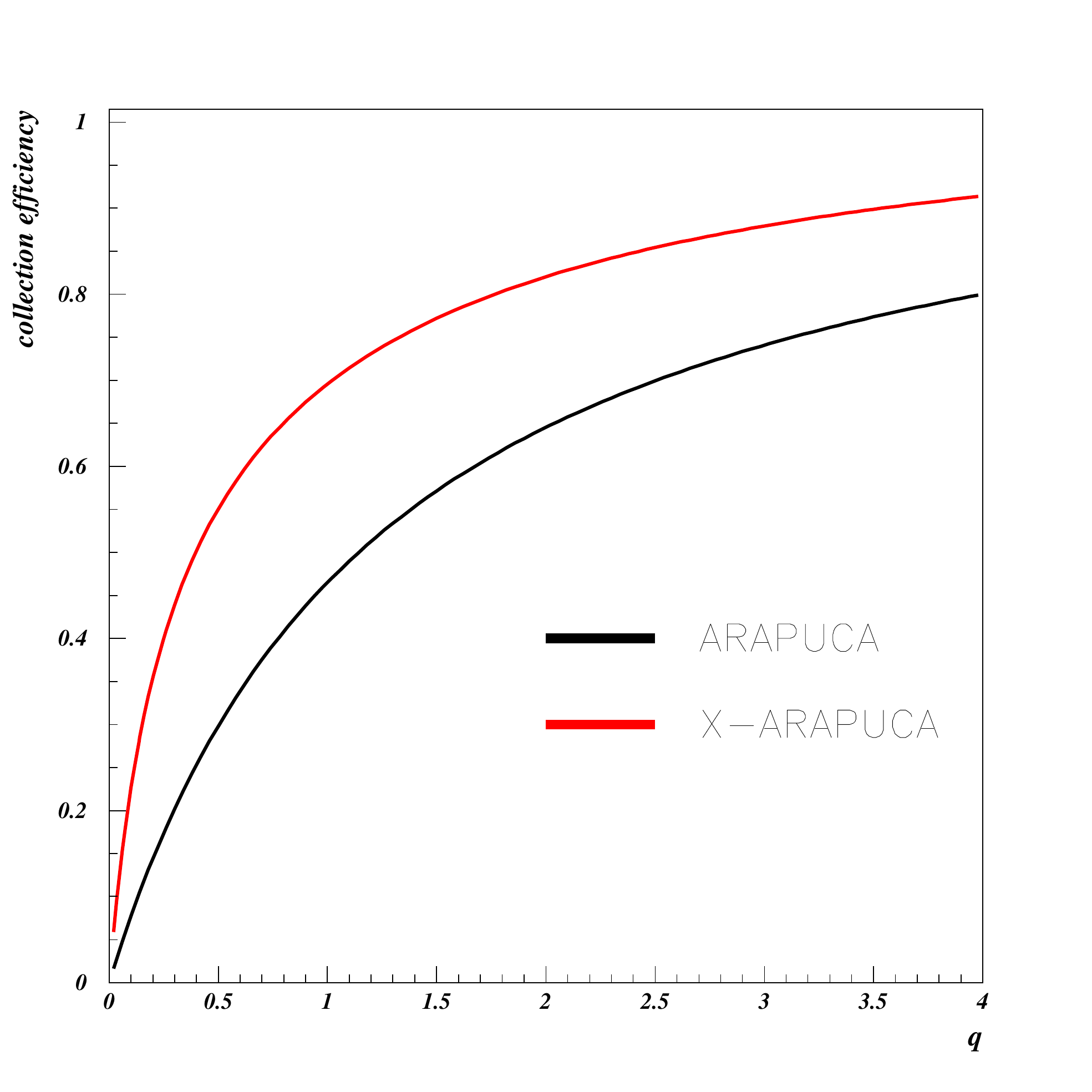}
	\includegraphics[width=7cm]{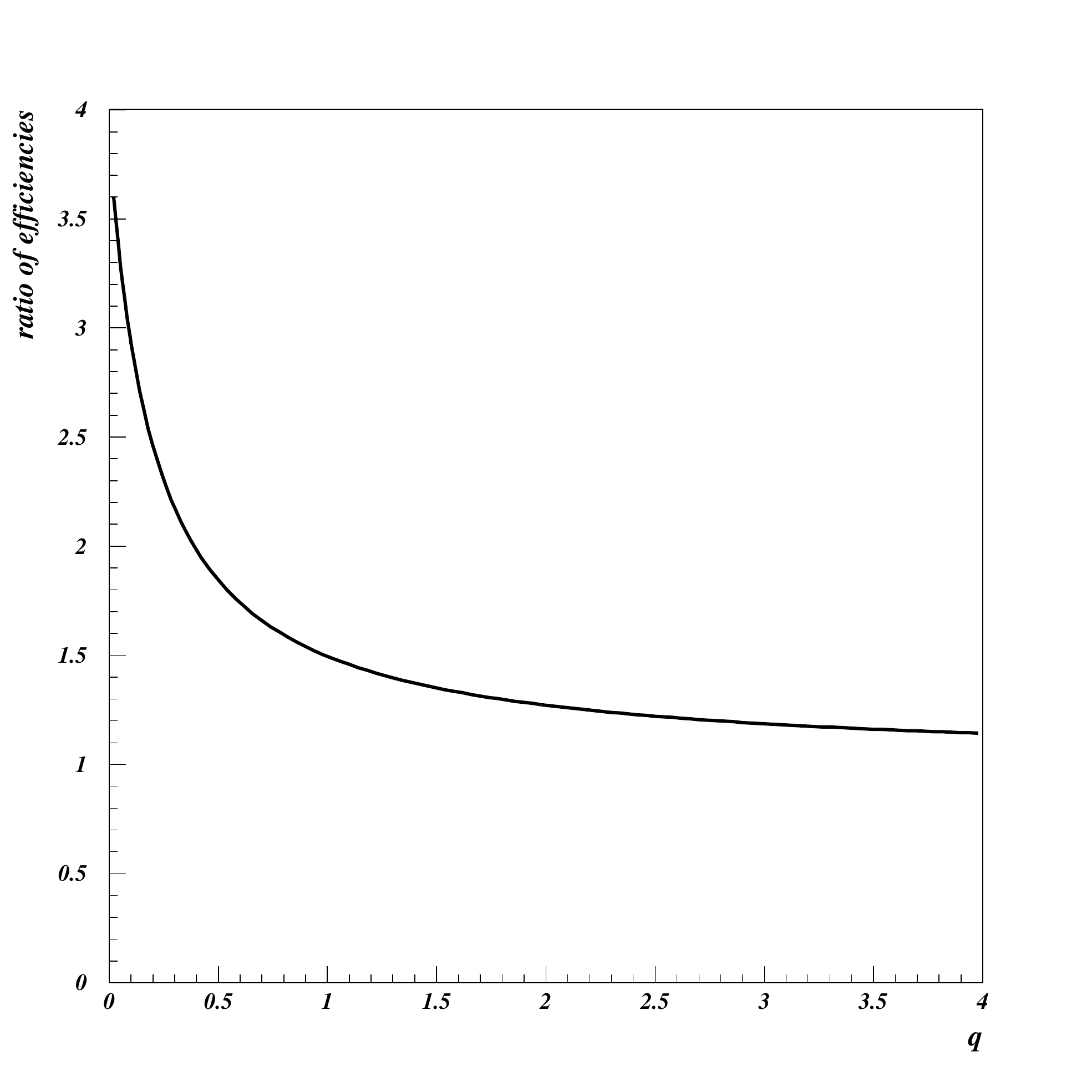}
	\caption{Left: collection efficiency of a square ARAPUCA and X-ARAPUCA as a function of the SiPM coverage of the lateral surface of the inner box. Zero corresponds to no coverage and four to total coverage. Right: ratio between the X-ARAPUCA and ARAPUCA collection efficiencies a a function of the SiPM coverage of the lateral surface of the inner box.  }
	\label{fig:collection_ratio}
\end{figure}

Setting the ratio {\it L/h} to 10 and considering an average reflectivity of 0.95\footnote{These represent typical values for the ARAPUCAs which are currently being built for the protoDUNE detector. In that case the box dimensions are 8$\times$10$\times$0.6 cm$^3$ and the average internal reflectivity is estimated to range between 0.93 and 0.95.}, equations \ref{eq:eff_coll_arapuca} and \ref{eq:eff_coll_xarapuca} can be plotted in terms of the parameter {\it q}, which is related to the active coverage of the lateral surface of the box. The plot is shown in figure \ref{fig:collection_ratio},left. The ratio between the collection efficiency of the X-ARAPUCA and that of the ARAPUCA is shown in figure \ref{fig:collection_ratio},right. 

 %The collection efficiency results to be $\epsilon_{coll}^{slab}$=0.95. The fraction of photons collected by the SiPM through total internal reflection inside the slab is then $\epsilon_{trapped} \times \epsilon_{coll}^{slab}$ $\simeq$ 0.54.

%The fraction of the photons which is not trapped inside the slab but is trapped by the standard ARAPUCA mechanism is (1-$\epsilon_{trapped}$)$\times$ $\epsilon_{coll}^{ARAPUCA}$ $\simeq$ 0.18.

%The total fraction of photons collect by the SiPM inside the X-ARAPUCA is then $\epsilon_{coll}^{X-ARAPUCA}$ $\simeq$ 0.54+0.18 = 0.72 

The collection efficiency of the X-ARAPUCA exceeds that of a standard ARAPUCA over the entire range of the parameter { \it q}. An interesting region is where {\it q} is less then one, where the increase in performances exceeds 50\%. 

%In this ideal case the effects of the absorption of the photons by the acrylic slab and of the losses of the photons due to imperfections in the slab planarity are neglected. In any case the latter effect can be compensated by the standard ARAPUCA trapping mechanism.

\section{Preliminary Studies Using a Monte Carlo Simulations}
A Monte Carlo (MC) simulation based on the Geant4 toolkit \cite{Agostinelli2003}
is being developed in order to perform a detailed study of the performances of the X-ARAPUCA and compare them to those of a standard ARAPUCA, taking into account all the possible mechanisms of photon production, propagation and detection.\\
The simulation currently propagates 350 nm or 430 nm photons inside the box, includes the absorption of the photons inside the acrylic slab, specular and diffusive reflections on the internal surfaces, refractive indexes of the materials, reflection/transmission at the boundaries. It does not include the optical properties of the dichroic filters, the emission spectra of the wavelength shifters and the SiPMs. 

% was used to perform a preliminary Monte Carlo (MC) study to compare the performances of the X-ARAPUCA with 

A preliminary MC study has been done to compare the collection efficiency of an X-ARAPUCA and of an ARAPUCA with the same active coverage of the internal surface with SiPMs.
%The preliminary MC study has been done with a 
%device which is the same of the one considered for the analytical calculation showed in section \ref{sec:analytical}.
The internal box is made of PTFE, it has dimensions of 80$\times$100$\times$10 mm$^3$ and is observed by an array of SiPM which covers an area of 80$\times$4 mm$^2$ and is installed on one of the lateral sides of the inner surface. 
The TPB doped acrylic light guide has a height of 0.4 cm and the absorption length for 430 nm photons is set conservatively at 50 cm. The average reflectivity of the internal surfaces is set to 0.98.
% The dichroic filter has the cutoff at 400 nm and dimensions of 8X10 $cm^2$. The shifter on the external side is para-Terphenyl (pTP) which shifts the light in a region around 350 nm (well below the filter cutoff) \cite{pTP} and its emission spectrum has been simulated.

\begin{figure}[h]
	\begin{center}
		\includegraphics[width=7cm,angle=0]{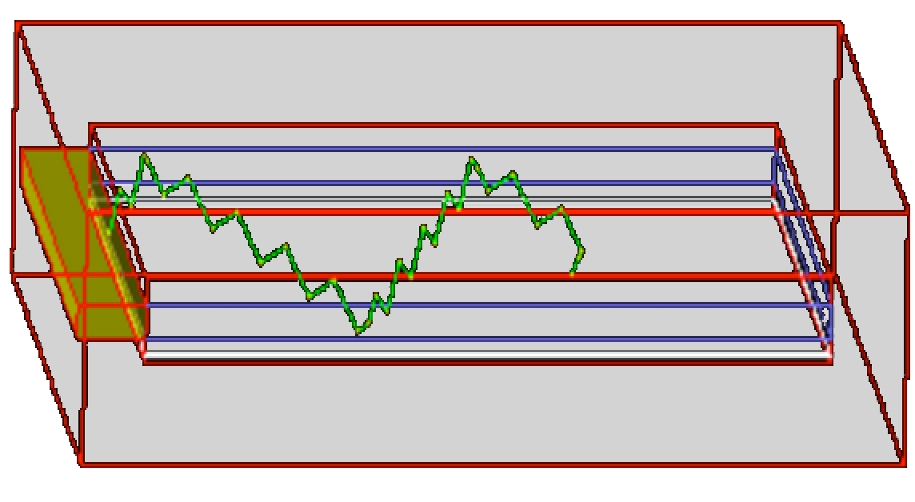}
		\includegraphics[width=7cm,angle=0]{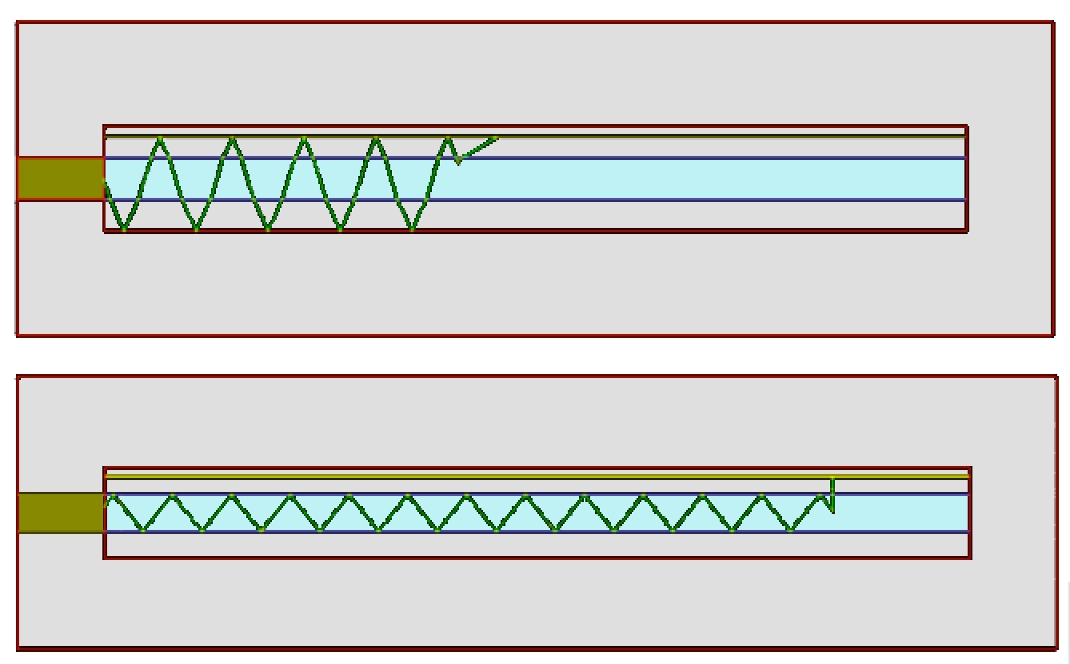}
		\caption{Left: pictorial representation of the device implemented in the Monte Carlo simulation; Right:two different trapping mechanisms are shown. A photon is trapped by standard ARAPUCA mechanism in the top right picture. A photon is trapped by total internal reflection inside the slab in the bottom right picture.}
		\label{fig:trapping}
	\end{center}
\end{figure}

In figure \ref{fig:trapping},left the simulated device is presented, while in figure \ref{fig:trapping},right two different trapping processes are shown.

In order to quantitatively evaluate the gain of the X-ARAPUCA with respect to the standard ARAPUCA, two simulations have been performed: one with the TPB doped guiding slab (X-ARAPUCA), and another without (ARAPUCA). 
%In the latter case the TPB is deposited on the inner side of th%e filter.
%The absorption length of the 430 nm photons inside the acrylic slab has been set to 50 cm.
The simulated trapping efficiency for the standard ARAPUCA was found to be around 38\%, while the inclusion of the guiding slab showed an increase in the efficiency of up to 53\%. The same collection efficiencies have been calculated with the analytical formulas \ref{eq:eff_coll_arapuca} and \ref{eq:eff_coll_xarapuca} and resulted to be $\sim$ 40\% and 70\% respectively. While the collection efficiencies for the ARAPUCA are compatible between each other, it seems that the analytical model tends to over-estimate the collection efficiency of the X-ARAPUCA. More investigations are ongoing in order to understand this discrepancy. In any case the advantage of the X-ARAPUCA over a standard ARAPUCA is still evident.   

%, which is less than what estimated analytically, but still significative. 
%Another advantage in using the X-ARAPUCA design is that photons with large angle of incidence (with respect to the slab surface) are guided by total internal reflections and only those with smaller angles are trapped through standard ARAPUCA mechanism. This allows to simplify the design and the performances of the dichroic filter in terms of reflectivity and transmissivity.   

\begin{figure}[h]
	\begin{center}
		\includegraphics[width=6cm,angle=0]{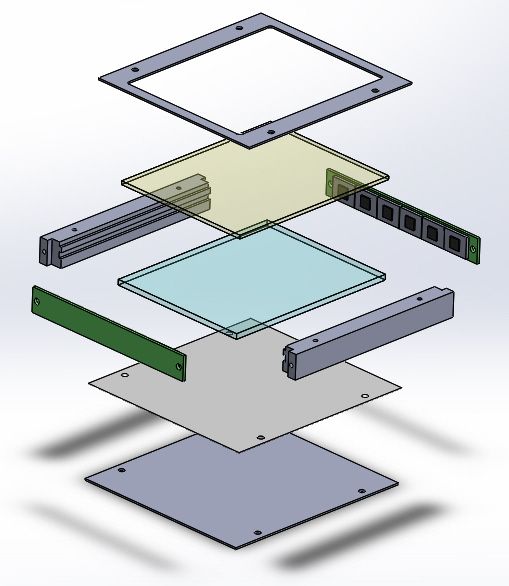}
		\caption{ Individual components of the X-ARAPUCA prototype. From top to bottom: filter holder, dichroic filter, acrylic slab doped with TPB, 3M VIKUITI foil, G-10 backplane. On two opposite lateral sides the two arrays of SiPM are installed, while the other two lateral  sides are made by G-10 lined with VIKUITI foils.}
		\label{fig:ARAPUCA_exploded}
	\end{center}
\end{figure}

\section{The first prototype}

The first X-ARAPUCA prototype is under construction at Colorado State 
University. The acceptance widow is a 8$\times$10 cm$^2$ dichroic filter 
with cutoff of 400 nm and coated with pTP on the external 
side. The light guide has the same dimensions of the acceptance 
window and is made by acrylic doped with TetraPhenyl-Butadiene (TPB) which 
has an emission wavelength centered around 430 nm. The bottom layer is a G10 
mechanical support backplane over which an ESR 3M VIKUITI \cite{VIKUITI} 
high reflective foil is installed. Also the lateral walls of the box will be 
build with G10 lined with VIKUITI foils. The active photo-sensors are 
Hamamatsu cryogenic SiPMs (active area of 6$\times$6 mm$^2$ each) of the 
same type of the ones which are being developed for the protoDUNE detector 
\cite{protoDUNE}. Two boards, holding 6 SiPM each, are installed on two 
opposite sides of the box and optically coupled to the acrylic slab.  

The dimensions of the filter and of the box are exactly the same as those 
used for the ARAPUCAs which are being installed in the protoDUNE detector. 
This will allow to partially re-use the same components and drawings. All 
the individual components of the X-ARAPUCA prototype are shown in figure \ref{fig:ARAPUCA_exploded}.\\

This first prototype will be tested in a LAr environment at the Universidade Estadual de Campinas (Brazil) and at the Universidade Federal do ABC (Brazil) in the next few months.

 %Two PCBs at each end hold Hamamatsu cryogenic SiPMs each, with active area of 6X6 mm$^2$ each.
 %which are the same dimensions of the ARAPUCAs installed in protoDUNE \cite{protoDUNE}. The latter is a large scale prototype of the DUNE far detector which is being constructed at CERN and which will be exposed to a charged particle beam in mid 2018. An array of ARAPUCAs will be operated inside this prototype. The acceptance widow is composed by a dichroic filter with a cutoff of 400nm and P-Therphenyl (350nm) on the external side. The dimension of the light guide is the same that the acceptance window. The bottom layer is a G10 mechanical support backplane, over this layer a 3M VIKUITI \cite{VIKUITI} high reflective foil will be installed. The  light guide will be made by an acrylic slab doped with TPB. The lateral walls will be build using G10 lined up with 3M VIKUITI foils. Two PCBs at each end hold 6 Hamamatsu cryogenic SiPMs each, with active area of 6X6 mm$^2$ each.

\section{Conclusions}
The X-ARAPUCA is a development of the original ARAPUCA concept, in which the combination of different well-established technologies allows to produce an even more efficient device. Preliminary studies based on analytical models and Monte Carlo simulations showed quite encouraging results, indicating that the collection efficiency of the X-ARAPUCA can be significantly higher than that of a conventional ARAPUCA.

\section{Acknowledgments}
This work is funded by FAPESP (Funda\c{c}\~ao de Amparo \`a Pesquisa do Estado de S\~ao Paulo) under the project 2016/01106-5 and supported by  
CNPq (Conselho Nacional de Desenvolvimento Cient\'ifico e Tecnol\'ogico).

\end{document}